\documentclass[preprint2]{aastex}

\usepackage{color}

\shorttitle{Comparison of photometric variability before and after stellar flares}
\shortauthors{C. Karoff}
\begin{document}
\title{Comparison of photometric variability before and after \\ stellar flares}
\author{C. Karoff}
\affil{Stellar Astrophysics Centre, Department of Physics and Astronomy, Aarhus University, Ny Munkegade 120, DK-8000 Aarhus C, Denmark\\and \\
Department of Geoscience, Aarhus University, H{\o}egh-Guldbergs Gade 2, 8000 Aarhus C, Denmark}
\email{karoff@phys.au.dk}

\begin{abstract}
The energy in the solar acoustic spectrum is known to be correlated with flares, but it is not known if the same is true for stellar flares? In order to answer this question, we have analyzed 73 flares in 39 solar-like stars. These flares were identified in the 854 solar-like stars observed by the $Kepler$ spacecraft that have stellar parameters measured with asteroseismology. Though we were not able to identify a statistically significant enhancement of the energy in the high-frequency part of the post-flare acoustic spectra compared to the pre-flare spectra of these stars, we did identify a larger variability between the energy in the high-frequency part of the post- and pre-flare acoustic spectra compared to spectra taken at random times.  
\end{abstract}


\section{Introduction}
Solar flares are large explosions on the solar surface that releases 10$^{29}$ to 10$^{32}$ erg of energy on a timescales from a few minutes to hours \citep{2008LRSP....5....1B}. The physical mechanism behind solar flares likely involves emergence of magnetic fields from the solar interior to the atmosphere and magnetic reconnection of these fields in the corona \citep{2011LRSP....8....6S}. Flares are known to occur on other stars too. The so-called super flares with energies between 10$^{33}$ to 10$^{38}$ erg have been seen on F--K main-sequence stars \citep[e.g.][]{2000ApJ...529.1026S, 2011AJ....141...50W, 2012Natur.485..478M}. Though these flares could be explained by a mechanism similar to the one operation in the Sun, it is harder to explain flares on M dwarfs, which are found to be able to increase the integrated flux from the stars with up to $\sim$200 times in the $U$-band \citep{2010ApJ...714L..98K}, on Mire Type stars \citep{1991ApJ...366L..39S}, which are giants and therefore only expected to have week magnetic fields, on RS CVn binaries \citep{1992MNRAS.255...48M} and recently on A stars, which are not expected to have a significant near-surface convection zone \citep{2012MNRAS.423.3420B}. 

If was first suggested by \citet{1972ApJ...176..833W} that large solar flare could stimulate free modes of oscillations to observable amplitudes. Later the detection of global low-degree p-mode oscillations was confirmed in observations of disk-interegrated sunlight \citep[by e.g.][]{1976Natur.259...92B, 1981Natur.293..443C} and soon it was generally accepted that the global p-mode oscillations were driven by stochastic excitation from granulation \citep{1988ApJ...326..462G} and not by solar flares, but significant effort has been investigated in detecting any correlation between flares and oscillations in the Sun -- here we will limit the discussion to the analysis of low degree or disk integrated observations \citep[for high-degree observations we referee to:][]{1988IAUS..123...59H, 1990SoPh..129...83B, 1998Natur.393..317K, 1999ApJ...513L.143D, 2005ApJ...630.1168D, 2007MNRAS.374.1155M, 2007ApJ...670L.147K, 2008SoPh..251..613M, 2010ApJ...710L.121R, 2011ApJ...734L..15K, 2011ApJ...739...70Z, 2011ApJ...741L..35Z, 2013SoPh..284..315Z}.

\citet{1999MNRAS.303L..63G} found a high correlation between temporally varying p-mode power measured at low degree in observations from the Global Oscillation Network Group (GONG) and coronal mass ejection event number, \citet{2006SoPh..238..219A} on the other hand found no correlation between a longer disk-integrated GONG data set, flares and coronal mass ejection indices. An analysis of observations from the Birmingham Solar Oscillations Network (BiSON) also revealed that though the strength of the p~modes follows the distribution expected from stochastic excitation by near-surface convection, there is evidence of a few more very large events than the numbers predicted by theory, but these events show only a poor correlation with flares \citep{1995soho....2..335C}. \citet{1998A&A...339..261F} and \citet{1998A&A...330..341F} made a statistical analysis similar to the one by \citet{1995soho....2..335C} and found a mean correlation between p~modes of degree 0 and 1 of 0.6 \% in observations from the Global Oscillations at Low Frequencies (GOLF) instrument on board the Solar and Heliospheric Observatory (SOHO) spacecraft  obtained from 1996-97 and 10.7 $\pm$ 5.9 \% in observations from the Interplanetary Helioseismology by Irradiance experiment (IPHIR) on the Phobos spacecraft obtained in 1988, which could suggest a relation between the p~modes and transient events; but no matches between the events in the p~modes and in the flares or coronal mass ejections were made. \citet{1992ApJ...394L..65B} analyzed the correlation between acoustic energy and activity, not in time, but in space and found that while the energy of the p~modes with frequencies between 2.5 and 4.0 mHz is suppressed in active regions the energy in the 5.5 to 7.5 mHz frequency range is enhances around the active regions.   

The discovery by \citet{2008ApJ...678L..73K} of a high correlation between the energy in the high-frequency part of the acoustic spectrum calculated from observations from the Variability of solar Irradiance and Gravity Oscillations (VIRGO) instrument on SOHO and the solar X-ray flux was therefore the first evidence that supported Wolffs idea. The reason why \citet{2008ApJ...678L..73K} were able to see a correlation where others had failed was that \citet{2008ApJ...678L..73K} did not look at the frequency range where most of the p-mode energy is positioned around 3 mHz. Instead they looked at frequencies higher than the atmospheric acoustic cut-off frequency (5.3 mHz).

\citet{2009ASPC..416..333K} analyzed the effect of the 28 October 2003 flare and found that the oscillation were significant enhanced during the flare in the high frequency band (5-6.5 mHz) of the acoustic spectrum from GONG observations. This study was followed up by a larger study using wavelet techniques on observations of major flare events in solar cycles 23 from both the Michelson and Doppler Imager (MDI) and the GOLF instrument both on board SOHO by \citet{2010ApJ...711L..12K}, who found clear indications of enhancements of the high-frequency part of the spectra from MDI, but only marginal indications of enhancements in the high-frequency part of spectra from GOLF . Recently, these studies have been extended by at study of the 13 December 2006 flare, which was followed by an enhancement in the high-frequency part of spectra from both GONG and GOLF \citep{2011ApJ...743...29K}.  

\citet{2011JPhCS.271a2048C} preformed an independent confirmation of the results by \citet{2008ApJ...678L..73K} using observations from both VIRGO and GONG. Here the correlation was found to be slightly smaller for the radial velocities calculated from GONG observations compared to intensity observations from VIRGO. The physical interpretation of different responses in velocity and intensity has been discussed by \citet{2009ASPC..416..233K}.

On there other hand \citet{2012SoPh..281...21R} found that a diminution was just as likely as an enhancement of the high-frequency part of the acoustic spectra taken after 31 flares, which led them to conclude that {\it flare-related variations are probably no different from the variations during quiet times that arise from the stochastic driving of the modes}. 

Though the super flares we observe on solar-like stars with i.e. the $Kepler$ spacecraft \citep{2011AJ....141...50W, 2012Natur.485..478M} might be fundamental different from the flares that we observe on the Sun, they still provide an opportunity to test the hypothesis that flares can excite oscillations in stars in general. Unfortunately, the analysis of other stars cannot be performed in the same way as it was done for the Sun. Mainly because the observations we have at hand, even from $Kepler$, does not have the same quality as the solar observations, but also because we do not have continuously X-ray observations of the flares on the other stars, as we have for the Sun. We have therefore analyzed the 854 stars, with oscillation excited to high enough amplitude to allow an asteroseismic analysis \citep{2011MNRAS.415.3539V} of the $Kepler$ observations\footnote{Originally, oscillations were only found in 642 out of 1948 F--K main-sequence stars using $Kepler$ observations from quarter 1 to 3, but this analysis was extended using all the observation form the survey phase \citep[including quarter 4, ][]{2010AN....331..972K}, which resulted in detection of oscillations in 854 out of  2641 F--K main-sequence stars \citep{2011ApJ...738L..28V}}. Using 37 months of observations from $Kepler$, we have identified 73 flares in 39 of these solar-like stars using the same technique that was used by \citet{2012Natur.485..478M} and made a simple comparison of the energy in the high-frequency part of acoustic spectra taken before and after the flares and of acoustic spectra taken at random time. 

\section{Observations}
For the analysis we used all available short cadence \citep[58.85 sec. sampling,][]{2010ApJ...713L.160G} observations from the $Kepler$ mission \citep{2010ApJ...713L..79K} of the 854 stars for quarter 1 to quarter 13 (18 June 2009 to 27 June 2012). As the aim of this study is to analyze the high-frequency part of the acoustic spectra, the short cadence observations were needed. In order to lower the risk of artifacts affecting the analysis, the observations were corrected with the Presearch Data Conditioning Maximum A Posteriori (PDC-MAP) algorithm \citep{2012PASP..124.1000S}. For most of the stars only one month of observations were available, but for a limited number of stars ($\sim$ 150) observations were available for up to 28 months. 

\section{Analysis}
Flares were identified in the light curves using the technique described by \citet{2012Natur.485..478M} -- i.e. flare candidates were found by calculating the distribution of brightness changes between all paris of two consecutive data points and then placing a flare candidate threshold on three times the value of the top 1\% of these distributions. The criteria used by \citet{2012MNRAS.423.3420B}, that no flares in any star were allowed to take place at the same time was also applied to the flare candidates and finally the duration and shape of the light curves were examined along with the pixel-level data before the flare candidates were confirmed as a flare events.

In this way 83 flares in 40 stars were detected. An example of a flare event is shown in Fig.~1. In fact two flares can be seen in this figure. The presence of two or more flares closely separated in time in the light curve is likely to influence the analysis of the photometric variability. The flare event shown in Fig.~1 was therefore withdrawn, along with 4 others, from the analysis -- leaving 73 flares in 39 stars for the rest of the analysis.

Though the flares were identified using the same technique that was used by \citet{2012Natur.485..478M} some of the flares identified in this study are likely smaller than the flares identified by \citet{2012Natur.485..478M} as the short cadence observations allow for detection of smaller flares. 

All the stars analyzed are shown in a Hertzsprung-Russell diagram in Fig.~2 with the 39 stars marked. It is seen that flares are found all over than main sequence. More surprising, it is also seen that 5 stars on the sub-giant branch show flares. This is unexpected as evolved stars are general believed to be slow rotators \citep{1972ApJ...171..565S} and thus not host strong magnetic fields \citep[see][for a discussion of the relation between flares and rotation]{2013arXiv1304.7361N}. 

The next step in the analysis was to calculate pre- and post-flare acoustic spectra. These spectra were calculated from substrings of different length before and after the flares. From all the time strings one hour before and after the flares were excluded from the analysis (indicated by the vatical lines in Fig.~1) in order to lower the contamination from transients in the analysis.

The acoustic spectra were calculated as least squares spectra \citep{1976Ap&SS..39..447L} and normalized by the effective length of a given substring. As it is not known on which time scale the seismic response to the flares, if any, take place the lengths of the substrings were varied between 1 and 80 hours.

\section{Results}
The photometric variability associated with the flares was evaluated by measuring the total energy in the high-frequency part of both the pre- and post-flare acoustic spectra. Here high-frequency means the region in which \citet{2008ApJ...678L..73K} saw the largest correlation -- i.e. the region around the atmospheric acoustic cut-off frequency (in the Sun this frequency is located around 5.3 mHz). For the analysis the high-frequency region was defined to be located between the frequency of maximum power plus 10 times the large frequency separation and the frequency of maximum power plus 20 times the large frequency separation. For the Sun this means the region between 4.5 and 5.8 mHz. The evaluation was done by calculating the distribution of the relative difference between the total energy in the high-frequency part of the pre- and post-flares acoustic spectra for all the 73 flares (the flare distributions). These distributions were then compared to the distributions of the relative difference between the total energy in the high-frequency part of two acoustic spectra taken at random times separated by 2 hours (as in the flares case) for the same stars (the random distributions). We then used a Kolmogorov-Smirnov test to evaluate if the two distributions were significant different. The Kolmogorov-Smirnov test calculates the distance between the empirical distribution functions of the two samples and uses this distance to calculate the probability that the two samples come from the same distribution. This evaluation was done using 80 different lengths of the substrings from 1 to 80 hours.

An example of a comparison of the two distributions is show in Fig.~3 for a substring of 3 hours - where the solid line is the flare distribution and the dashed line is the random distribution. The two distributions are clearly difference, which is also revealed by the Kolmogorov-Smirnow test. In general all the distributions calculated using substring lengths shorter than 30 hours were significant different, this can be seen in Fig.~4 that shows the p value from the Kolmogorov-Smirnow test as a function of the length of the substrings. Here it is also seen that for substrings longer than 45 hours the distributions are generally not significant different.

On the other hand Fig.~3 also shows that there is not a significant general enhancement in the high-frequency part of the post-flare acoustic spectra compared to the pre-flare acoustic spectra. The same was the case for the other substring lengths -- i.e. that the mean values of the flare distributions were not significant different from zero. In other words do the $Kepler$ observations not show a significant enhancement of the photometric variability at high frequency at times of flares in the 39 solar-like stars analyzed in this study.

In Fig.~3 it can also be seen that the significant difference between the flare distribution and the random distribution is cause by a larger width and different shape of the flare distribution compared to the random distribution. In order to evaluate if this was a general phenomena for all substring lengths, we measured the widths of the flare distributions (calculated as the variance of the relative difference between the total energy of the high-frequency part of the acoustic spectra). Fig.~5 shows these widths as a function of the length of the substrings. Here the solid line represents the width of the flare distributions whereas the dashed line represents the width of the random distributions. It is seen that the widths of the flare distributions are significant larger than the widths of the random distributions for substring lengths shorter than 45 hours. 
It can therefore be concluded that the photometric variability at high frequency of solar-like stars is significant modified at times of flares, but the modification is equally likely to be a diminution as an enhancement.

In order to test the robustness of this conclusion, the analysis was redone using slightly different definitions of the high-frequency region and we also tried to measure the relative energy enhancement between a high-frequency and a low-frequency region, instead of just the energy enhancement in the high-frequency region. All these tests confirmed that the conclusion above were robust against changing these parameters. 

The modulation of the photometric variability could be caused by additional smaller flares located in the light curves either before or after the main flare events. In order to eliminate this possibility, a Kolmogorov-Smirnov test was applied to compare the distributions of the numbers of such tiny photometric events either before or after the main flare events to random cases. These tiny photometric events were identified in a way similar to how the flares were identified using the brightness change between all paris of two consecutive data points in the 80 different substrings of the 73 flares, but here the threshold was lowered to just 4$\sigma$. This returned a mean p value of 0.53 and 0.45 for the tiny events found before and after the main flare events, respectively. The distributions of the amplitudes of the highest change in brightness were also compared in the same way, which returned a p value of 0.43 for the tiny events found both before and after the main flare events. No dependency between the lengths of the substrings and the p values was found in any of the tests. It is therefore safe to conclude that the modulation of the photometric variability is not caused by additional smaller flares located in the light curves either before or after the main flare events.

\section{Discussion} 
It seems likely that the super flares we observe on other solar-like stars are caused by a physical mechanism similar to the one operating in the Sun \citep{2013arXiv1304.7361N} and therefore it also seem likely that flares would affect the oscillations in other stars in a way similar to how they seem to affect the oscillations in the Sun \citep{2008ApJ...678L..73K}. On the other hard super flares can be up to 6 order of magnitudes stronger that the strongest flares we have observed on the Sun \citep{2000ApJ...529.1026S}, so the scenario could also be rather different.

Using observations from $Kepler$ of 73 fares on 39 solar-like stars it was possible to concluded that the photometric variability of solar-like stars is significant modified at times of the flares. If this is indeed caused by a seismic response to the flares it provides a very strong argument in favour of the interpretation that the correlation between flares and the energy in the high-frequency part of the acoustic spectrum of the Sun found by \citet{2008ApJ...678L..73K} is due to a causal relation between flares and global oscillations. On the other hand it is not possible to eliminate other scenarios. 

The analysis also revealed that the modification of the photometric variability at high frequency of solar-like stars at times of flares was equally likely to be a diminution as an enhancement. This could point to some, hitherto unknown, mechanism in the flares that can lead to either excitation or damping of the global oscillations.

\acknowledgments
I thank the referee for thoughtful comments, which significantly improved the paper. This work was partially supported by NASA grant NNX13AC44G. Funding for this Discovery mission is provided by NASAs Science Mission Directorate. I also acknowledged support from the Carlsberg and Villum foundations. Funding for the Stellar Astrophysics Centre is provided by The Danish National Research Foundation (Grant agreement no.: DNRF106). The research is supported by the ASTERISK project (ASTERoseismic Investigations with SONG and Kepler) funded by the European Research Council (Grant agreement no.: 267864).

\begin{figure}
\plotone{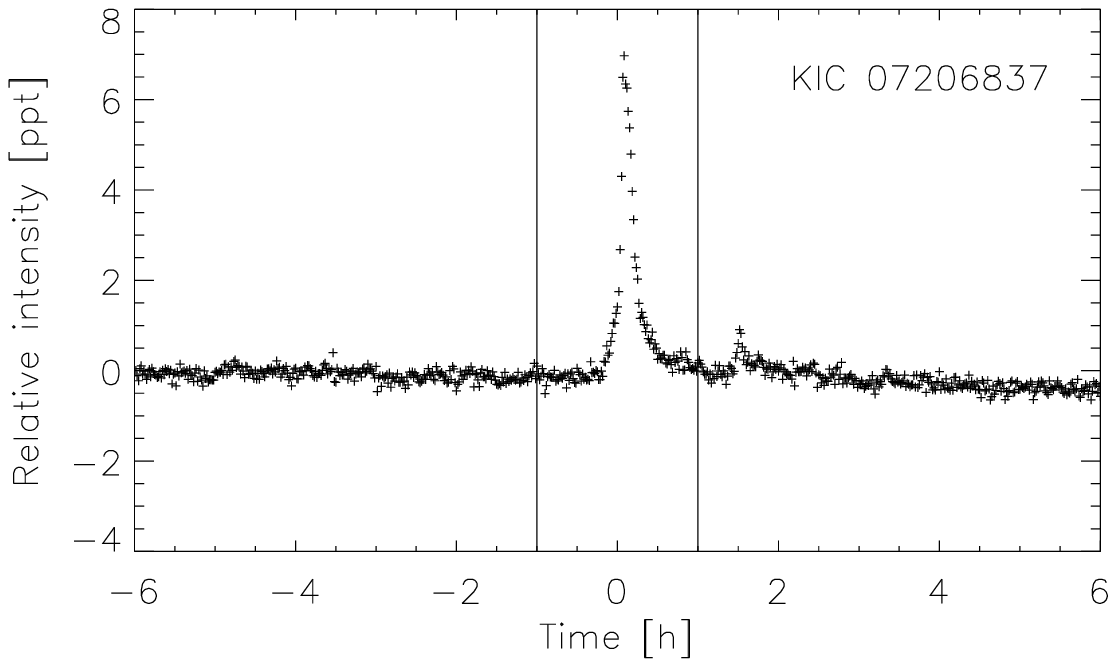}
\caption{Example of a flare in a short cadence $Kepler$ light curve. The vertical lines mark the $\pm$ 1 h region excluded from the analysis in order to lower the contamination from transients in the analysis. This flare event was excluded from the analysis along with 4 other flare events as the light curve does in fact show two flares separated by a little less than two hours.}
\end{figure}

\begin{figure}
\plotone{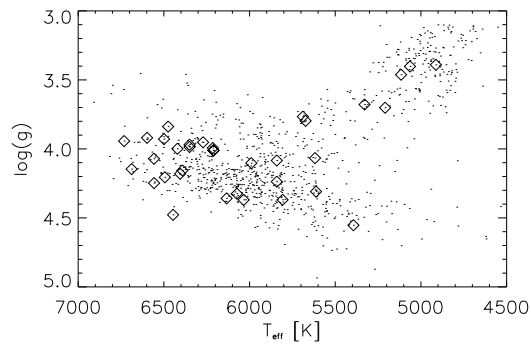}
\caption{Hertzsprung-Russell diagram of all the stars analyzed in this study, with the 39 flare stars marked. It is seen that flares are found on 5 stars on the sub-giant branch. The surface gravities and effective temperatures are from \citet{2011ApJ...738L..28V}.}
\end{figure}

\begin{figure}
\plotone{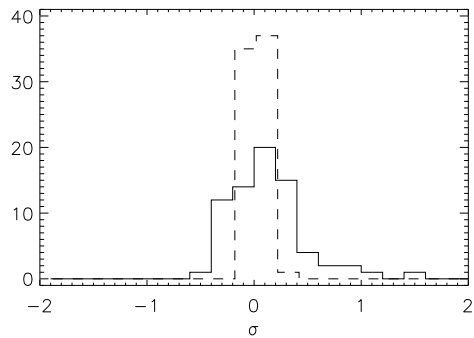}
\caption{Example of a histogram of the distribution of the relative difference between the energy in the high-frequency part of the pre- and post-flare acoustic spectra for the 73 flares in the 39 stars (solid line), compared to the distribution of the relative difference between the energy in the high-frequency part of two acoustic spectra taken at random times separated by 2 hours (as in the flare case) for the same stars (dashed line). The mean value of the relative difference between the energy in the high-frequency part of the pre- and post-flare acoustic spectra is seen to be close to zero, but the width of the flare distribution is, on the other hand, seen to be significant larger than the width of the random distribution. }
\end{figure}

\begin{figure}
\plotone{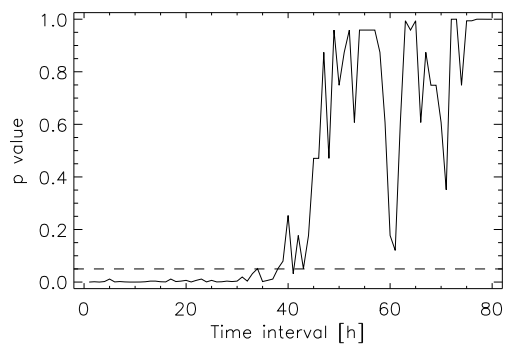}
\caption{p values from the Kolmogorov-Smirnow test as a function of the length of the substrings. It is seen that all the distributions calculated using substring lengths shorter than 25 hours are significant different. The vertical dashed line shows a 5\% confident level.}
\end{figure}

\begin{figure}
\plotone{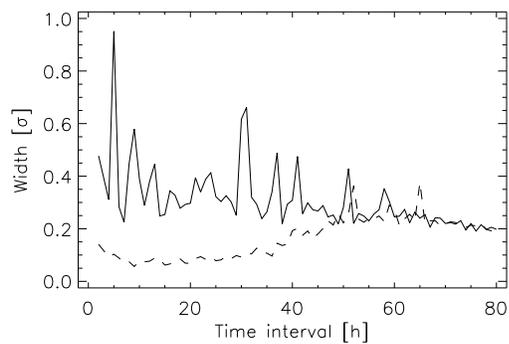}
\caption{The width of the distributions as a function of the length of the substrings. The solid line shows the width of the flare distributions whereas the dashed line shows the width of the random distributions. The larger width of the flare distributions compared to the random distributions suggest that the photometric variability of solar-like stars is significant modified at the times of flares.
}
\end{figure}

\end{document}